\documentclass{jnmp}

\usepackage{amsmath}
\JNMPnumberwithin{equation}{section}

\newtheorem{theorem}{Theorem}
\newtheorem{lemma}{Lemma}
\newtheorem{corollary}{Corollary}

\theoremstyle{definition}
\newtheorem{definition}{Definition}
\newtheorem{remark}{Remark}

\begin{document}

\renewcommand{\evenhead}{V~N~Grebenev and B~B~Ilyushin}
\renewcommand{\oddhead}{Invariant Sets and Explicit Solutions}

\thispagestyle{empty}

\FirstPageHead{9}{2}{2002}{\pageref{grebenev-firstpage}--\pageref
{grebenev-lastpage}}{Article}

\copyrightnote{2002}{V~N~Grebenev and B~B~Ilyushin}

\Name{Invariant Sets and Explicit Solutions\\
 to a  Third-Order Model
for the Shearless Stratified Turbulent
Flow}\label{grebenev-firstpage}

\Author{V~N~GREBENEV~$^\dag$ and B~B~ILYUSHIN~$^\ddag$}

\Address{$^\dag$~Institute of Computational Technologies,
Lavrentjev ave.~6, Novosibirsk 630090, Russia \\
~~E-mail: vova@lchd.ict.nsc.ru\\[10pt]
$^\ddag$ Institute of Thermophysics, Lavrentjev ave.~1,
Novosibirsk 630090, Russia\\
~~E-mail: ilyushin@itp.nsc.ru}

\Date{Received June 24, 2001; Revised October 21, 2001; Accepted
October 22, 2001}

\begin{abstract}
\noindent
We study dynamics of the shearless stratified
turbulent flows.
Using the method of differential constraints we find a class of
explicit solutions
to the problem under consideration and establish that the
differential
constraint obtained coincides
with the well-known Zeman--Lumley model for stratified flows.
\end{abstract}

\section{Introduction}
The results of this article continue our earlier studies
of the problem of interaction and~mi\-xing
between two semi-infinite turbulent flow fields of different
scales given in~\cite{GI,IG} wherein it was proposed
a concept based on the method of differential constraints
\cite{YA,SID} for exa\-mining the closure procedure for
momentum equations in Parametric Turbulent Models.
The key idea of this approach can be formulated shortly as
follows:
the algebraic expressions for the $n$-order moments of
statistical
characteristics of turbulent flow are deter\-mined as the
equations
of
invariant sets (manifolds) of the corresponding differen\-tial
equations.
As an illustrative example we have shown how this concept can be
applied~for finding a~selfsimilar solution to the
above-mentioned problem in the case of nonstratified~flow.

The aim of the present article is to extend the proposed method
to
the general case including stable and unstable stratification of
the flow. At first, we find a differential constraint
generated by the model under consideration and
then we construct a reduction that enables us to
rewrite that model in a more simple form on the
invariant set obtained both for stable and unstable
stratifications. This makes it possible to find explicit
solutions. The analysis of those solutions shows that the
influence of stratification on statistical characteristics of
turbulent flow is essential. The time scales of turbulence for
the all cases are founded in explicit forms and we show that
their behaviors are different. As an application to the Theory of
Parametric Turbulent Models, we indicate that the
differential constraint obtained coincides with the well-known
Zeman--Lumley model~\cite{ZEM}.

The motivation for the study is that, in many
practical
situations, a turbulent flow is embedded in a surrounding field
of
turbulence of different intensity. The shearless turbulent flows
appears in decaying grid turbulence in which the mean velocity
is constant throughout. Those flows include both the
homogeneously-damped grid turbulence and shearless turbulent
mixing layer which is formed beyond a composite grid.

\section{The governing equations}
The directions of turbulent-flow investigations based on
constructing closed equations of turbulent transport are used
intensely in modelling. The approach based on two-parametric
models of turbulence gained wide application, as well as one
based
on second-order closure models (for example, the $K-\epsilon$
model), which are effective from the computational viewpoint and
yields results whose accuracy is sufficient for practical
applications. That models are based on parametrization of
third-order moments of the gradient type. However, the use of
that
models for description of turbulent transport in stratified flows
gives a qualitatively incorrect result (in some cases) see, for
example~\cite{LAM}. The anisotropic character of the buoyancy
effect on the structure of turbulence is manifested in appearance
of the long wavelength spectrum of turbulent
oscillations~\cite{MY}. This spectral range corresponds to
large-scale vortex structures containing the main portion of
turbulent energy. According to experimental and theoretical
studies, the following large-scale vortex structure is formed in
stratified flows: turbulent spots in the case of stable
stratification and coherent structures in the case of unstable
stratification which are mainly responsible for turbulent
transport. The effect of intermediacy and asymmetry of vertical
turbulent transport caused by the influence of large-scale vortex
structures makes the probability distributions of turbulent
fluctuations significantly non-Gaussian. The turbulent structure
in these flows is usually described by third-order closure models
where the triple correlations (asymmetry) are calculated from
differential transport equations. There is considerable number of
references connected with the use of third-order turbulent
mo\-dels. However, as a rule these models employ
Millionshchikov's
quasinormality hypothesis for the parametrization of diffusion
processes in equations for triple correlations and, according to
this hypothesis, all cumulants of fourth- and higher-order may be
negligibly small in comparison with the corresponding correlation
functions.  As a consequence, in the former case the
triple-correlation equations are of the first-order without a
dumping mechanism for triple correlations that leads to
physically
contradictory results~\cite{LAM}. The approach proposed in
\cite{IL1,IL2} allows us to overcome this obstacle; the technique
also includes a physically reasonable way for constructing
approximate algebraic parametrizations of higher moments. Using
these observations, we present a third-order model of turbulence
to describe correctly the shearless turbulent mixing layer in the
framework stated above.

To determine the values of the horizontal components
$e_h$ of the turbulent kinetic energy
$e = e_h + 1/2\langle w^2 \rangle$ and
the rate of
dissipation $\epsilon$ and the one-point velocity correlation
$\langle w^2\rangle$ of the second-order, we make use of the
differential equations:
\begin{gather*}
\frac{\partial e_h}{\partial t} = -
\frac{\partial\langle e_hw\rangle} {\partial z} -
\frac{c_1}{\tau}\left[e_h - \frac{2}{3}E\right]-
\frac{2}{3}\epsilon,
\end{gather*}
\begin{gather*}
 \frac{\partial\epsilon}{\partial t} =
\frac{\partial} {\partial z}\left[c_d\tau \langle
w^2\rangle\frac{\partial\epsilon}{\partial z}\right] +
 \frac{c_{\epsilon_1}}{\tau}\beta g\langle w\theta\rangle -
c_{\epsilon_2}\frac{\epsilon}{\tau},
\\
\frac{\partial \langle w^2\rangle}{\partial t}  =
- \frac{\partial
\langle w^3\rangle}
{\partial z} + 2\beta g\langle w\theta\rangle
- \frac{c_1}{\tau}
\left[\langle w^2\rangle - \frac{2}{3}E\right]-
\frac{2}{3}\epsilon.
\end{gather*}
Here $E$, $\tau = E/\epsilon$ are the kinetic energy and the
time scale of turbulence respectively. The volumetric expansion
coefficient is $\beta = 1/\Theta$; $\Theta$ and $\theta$
are the mean and variance potential temperatures respectively.
The constants involved in the model with the lower case letters
are denoted by $c_{**}$. The system of equations is
nonclosed and we complete it by the transport equation
for the triple correlation of the vertical velocity fluctuation:
\[
\frac{\partial\langle w^3\rangle}{\partial t}  = -\frac{\partial
C}
{\partial z} - 3\langle w^2\rangle\frac{\partial\langle
w^2\rangle}
{\partial z}
+ 3\beta g\langle w^2\theta\rangle - c_2\frac
{\langle w^3\rangle}{\tau},
\]
where $C = \langle w^3\rangle - 3\langle w^2\rangle^2$ is the
cumulant of velocity fluctuations.
To obtain a closed model of turbulent transport that does not
imply equality to zero of the fourth-order cumulants,
the closure procedure is performed at the level of the fifth
moments, i.e.\ it is assumed that the fifth-order cumulants are
equal to zero. Thus, to obtain the distribution of
the third-order cumulants (correlations), the latter are
calculated from differential transport equations:
the fourth-order cumulants are determined approximately
(from algebraic exp\-ressions), and the fifth-order cumulants
are assumed to be equal to zero, since their contribution is
negligibly small. Results~\cite{KOS} of numerical simulation
of vertical turbulent transport in a convective boundary layer
confirm the validity of this approach.
To express the fourth-order cumulant~$C$, we can use the equality
\[
C  = -\frac{\tau}{c_3}\left[6\langle
w^3\rangle\frac{\partial\langle w^2\rangle} {\partial z} +
4\langle w^2\rangle\frac{\partial\langle w^3\rangle} {\partial
z}\right].
\]
The triple correlation $\langle w^2\theta\rangle $, the vertical
heat flux $\langle
w\theta\rangle$ and the temperature
dispersion~$\langle\theta\rangle$
 are approximated by the following
algebraic
relationships~\cite{IL2}:
\begin{gather*}
\langle w^2\theta\rangle  =
-\frac{\tau}{c_4}\left[\langle w^3\rangle
\frac{\partial\Theta}{\partial z} - 2\beta g\langle
w\theta^2\rangle\right],
\qquad
\langle w\theta^2\rangle  =
-\frac{\tau}{c_5}\langle
w^2\rangle\frac{\partial\langle\theta^2\rangle}
{\partial z},
\\
\langle w\theta\rangle  =
-\frac{\tau}{c_{\theta_1}}\langle
w^2\rangle\frac{\partial\Theta}{\partial z}
\equiv - \frac{\tau N^2}{\beta gc_{\theta_1}}\langle w^2\rangle,
\qquad
N^2 = \beta g\frac{\partial\Theta}{\partial z},
\\
\langle \theta^2\rangle  =
-\frac{\tau}{c_{\theta_1}r}\langle
w\theta\rangle\frac{\partial\Theta}{\partial z}=
 - \frac{\tau N^2}{\beta gr}\langle w\theta\rangle
= \left(\frac{\tau
 N^2}
{\beta g}\right)^2\frac{\langle w^2\rangle}{c_{\theta_1}r},
\end{gather*}
where
$r=\tau/\tau_{\theta}$, $\tau_{\theta}$ is the time scale of
potential
temperature variance, $N$ is the Brunt--Vaisala frequency.
On using the balance approximation between exchange mechanism and
dissipation, the equation for the
horizontal component $e_h$ of the turbulent kinetic energy may be
simplified and may be written as follows:
\[
-c_1\left[e_h - \frac{2}{3}\left(e_h + \frac{\langle
w^2\rangle}{2}\right)
\right] = \frac{2}{3}\left(e_h + \frac{\langle
w^2\rangle}{2}\right).
\]
Hence,
\[
e_h =  \frac{c_1 - 1}{c_1 +2}\langle w^2\rangle,\qquad
E = \frac{3c_1}{2(c_1 + 2)}\langle w^2\rangle,\qquad
\tau = \frac{3c_1}{2(c_1 + 2)}\frac{\langle
w^2\rangle}{\epsilon}.
\]
As a result of the above simplifications, the model of the
shearless mixing turbulence layer of third-order
is represented by the following system of differential
equations
\begin{gather*}
\frac{\partial\langle w^2\rangle}{\partial t}  =
- \frac{\partial\langle w^3\rangle}{\partial z} - \frac{c_1}{c_1
 + 2}\frac{\langle w^2\rangle}{\tau^*},
\\
\frac{\partial \langle w^3\rangle}{\partial t}  =
\frac{\partial} {\partial z}\left[
\frac{\tau}{c_3}\left(6\langle w^3\rangle\frac{\partial\langle
w^2\rangle}
{\partial z} + 4\langle w^2\rangle\frac{\langle\partial
w^3\rangle}
{\partial z}\right)\right]
 - 3\langle w^2\rangle\frac{\partial\langle w^2\rangle}
{\partial z}  - c_2\frac{\langle w^3
\rangle}{\tau^{**}},
\\
\frac{\partial\epsilon}{\partial t}  =  \frac{\partial}
{\partial z}\left[c_d\tau
\langle w^2\rangle\frac{\partial\epsilon}{\partial z}\right]
 - c_{\epsilon_2}\frac{\epsilon}{\tau^{***}}.
\end{gather*}
Here
\[
\tau^{*} \approx \frac{\tau}{1 + \frac{6(c_1 + 2)}
{(c_1 + 4)c_{\theta_1}\tau^2N^2}},\qquad
\tau^{**} \approx \frac{\tau}{1 +
\frac{2}{c_2c_4\tau^2N^2}},\qquad
\tau^{***} \approx \frac{\tau}{1 + \frac{c_{\varepsilon_1}(c_1 +
c_2)}
{c_{\varepsilon_2}(c_1 + 1)c_{\theta_1}\tau^2N^2}}.
\]
It follows from the formula for $C$ \cite{IL2} that the
contribution of
the second term in the algebraic model for the cumulant $C$ is
essential. Also, we can assume that (see~\cite{WE})
\[
\tau^* = \tau^{**} = \tau^{***} =
\frac{\tau}{1 + \frac{\pi}{18}\tau^2N^2} = \tau_w.
\]
Thus the governing equations are:
\begin{gather} \label{1.1}
\frac{\partial\langle w^2\rangle}{\partial t}  =
- \frac{\partial\langle w^3\rangle}{\partial z}
- \alpha\left(1 + a\hat\tau^2N^2\right)
\frac{\langle w^2\rangle}
{\hat\tau},
\\ \label{1.2}
\frac{\partial \langle w^3\rangle}{\partial t}  =
\frac{\partial} {\partial z}\left[ \kappa\hat\tau \langle
w^2\rangle\frac{\partial\langle w^3\rangle}{\partial z} \right]
- 3\langle w^2\rangle\frac{\partial\langle
 w^2\rangle} {\partial z}  - \gamma\left(1 +
 a\hat\tau^2N^2\right)
\frac{\langle w^3\rangle}{\hat\tau},
\\ \label{1.3}
\frac{\partial\epsilon}{\partial t}  =  \frac{\partial}
{\partial z}\left[\delta\hat\tau
\langle w^2\rangle\frac{\partial\epsilon}{\partial z}\right]
 - \varrho\left(1 +
 a\hat\tau^2N^2\right)\frac{\epsilon}{\hat\tau},
\end{gather}
where
$\hat\tau = \langle w^2\rangle/\epsilon$ and
$\alpha = 2/3$, $\kappa = 6c_1/c_3(c_1 + 2)$,
$\gamma = 2c_2(c_1+1)/3c_1$,
$\delta = 3c_1c_d/2(c_1 + 2)$,
$\varrho = 2c_{\epsilon_2}(c_1 + 2)/3c_1$,
$a = c_1^2\pi/18(c_1 + 2)^2$.

In addition, we indicate the equation for $\hat{\tau}$
\begin{gather}
\frac{\partial\hat\tau}{\partial t} =
- \frac{\hat\tau}{\langle w^2 \rangle}
\left[\frac{\partial\langle w^3\rangle}{\partial z}
+\delta\hat\tau\langle w^2\rangle\frac{\partial^2\langle
w^2\rangle}{\partial z^2} + \delta\hat\tau
\left(\frac{\partial\langle w^2\rangle}{\partial
z}\right)^2\right]
\nonumber\\
 \label{7}
\phantom{\frac{\partial\hat\tau}{\partial t} = }{} +
\delta\langle w^2\rangle\hat\tau\frac{\partial^2
 \hat\tau}{\partial
z^2}  + 2\delta\hat\tau\frac{\partial\hat\tau}{\partial
z}\frac{\partial\langle w^2
\rangle}{\partial z} -
\delta\left(\frac{\partial\hat\tau}{\partial z}\right)^2\langle
w^2\rangle +
(\varrho - \alpha)\left(1 + a\hat\tau^2N^2\right)
\end{gather}
which can be obtained from (\ref{1.1}), (\ref{1.3}).
This equation will be crucial for the further study of
properties of system (\ref{1.1})--(\ref{1.3}).

\section{Invariant sets}

In this section we show that system (\ref{1.1})--(\ref{1.3})
admits an invariant set (manifold) of the form
\begin{equation} \label{H1}
D =  \left\{\langle w^2\rangle, \langle w^3\rangle, \hat\tau\,
:\,
{\cal H}^1(\langle w^2\rangle, \langle w^3\rangle, \hat\tau)
\equiv \langle w^3 \rangle + \delta\hat\tau\langle
w^2\rangle\langle w^2\rangle_z = 0\right\}.
\end{equation}
A possible application of the existence of an invariant set
(manifold) generated by the differential equation for the
triple correlation of the vertical velocity fluctuations consists
in
constructing explicit solutions to the model obtained. Here
``explicit''
means solutions which can be found by means of ordinary
differential
equations (or algebraic systems).

Let us briefly present the special terminology of Symmetry
Analysis
(see
\cite{IBR,KAP} for more details).

Consider a system of evolution equations ${\cal F}$
\[
u_t^i = {\cal F}^i\left(t, x_1, \dots, x_n, u^1,\dots,
u^k_{\lambda},\dots\right)
\]
where $i=1,\dots, m$, $u_{\lambda}^k = \partial^{\lambda} u^k/
\partial x_1^{\lambda_1}\cdots\partial x_n^{\lambda_n}$.

A set(manifold) ${\cal H}$ given by equations
\[
h_i\left(t, x_1,\dots, x_n,\dots, u^1,\dots, u^m, \dots,
u_{\lambda}^k,\dots\right) = 0
\]
is said to be the invariant set (manifold) of system ${\cal
F}$ if
\begin{gather*}
V_{\cal F}(h^i)\left|_{[{\cal H}]_0}\right. = 0,
\\
V_{{\cal F}} = \frac{\partial}{\partial t} + \sum_{i=1}^m{\cal
F}^i
\frac{\partial}{\partial u^1} +
\sum_{i=1}^{m}D^{\alpha}({\cal F}^i)\frac{\partial}{\partial
u^i_{\alpha}},
\end{gather*}
where $\alpha = (\alpha _1,\dots,\alpha _n)$, $D^{\alpha} =
D^{\alpha _1}_{x_1}
\cdots D^{\alpha _n}_{x_n}$.

The invariant condition can be written in the following
equivalent form
\[
D_t (h_i)\left|_{[{\cal F}]_0}\right.\left|_{[{\cal
H}]_0}\right.= 0.
\]
Here $[{\cal F}]_0$ is $\infty$-prolongation (see \cite{KAP}) of
${\cal F}$
with respect to $x_1,\dots,x_n$. The set $[{\cal H}]_0$ is
determined by analogy.

To show that $D$ is invariant under the flow generated
by equation (\ref{1.2}), we prove that the operator ${\cal
H}^1(\langle w^2\rangle,
\langle w^3\rangle, \hat\tau) \equiv \langle w^3 \rangle +
\delta\hat\tau\langle w^2\rangle\langle w^2\rangle_z$ preservs
the sign on the
set of sufficiently smooth solutions to system
(\ref{1.1})--(\ref{1.3}).

We introduce into consideration the following sets
\begin{gather}
\label{H2}
D^+ =\left\{p, q, w\in C^{\infty}({\mathbb R})\,:\,
{\cal H}^1(p,q,w) \geq 0\right\},\\
\label{H3'}
D^- = \left\{p, q, w\in C^{\infty}({\mathbb R})\,:\,
{\cal H}^1(p,q, w)\leq 0\right\}.
\end{gather}

\begin{definition}
Operator ${\cal H}^1$ is said to be sign-invariant of
system
(\ref{1.1})--(\ref{1.3}) if
\begin{gather*}
\left(\langle w^2(\cdot,t_0)\rangle, \langle
w^3(\cdot,t_0)\rangle,
\hat\tau(\cdot,t_0)\right) \in D^+(D^-) \\
\qquad \Rightarrow  \
\left(\langle w^2(\cdot,t)\rangle, \langle w^3(\cdot,t)\rangle,
\hat\tau(\cdot,t)\right)
\in D^+(D^-), \quad t > t_0.
\end{gather*}
\end{definition}

\begin{remark}For quasilinear parabolic equations derivation of
suitable
sign-invariants
plays an important role for existence, uniqueness, regularity,
existence
of explicit solutions and other problems (see, for example,
\cite{KAL,GAL}).
\end{remark}

It is convenient to determine the invariant set $D$ under the
flow
generated by equation~(\ref{1.2}) via the sign-invariant
sets $D^+$ and $D^-$ as their intersection. The invariance
condition
takes the form
\[
\left.\frac{\partial}{\partial t}{\cal H}^1\right|_{{\cal H}^1 =
0}= 0.
\]
As the first result that uses
the above notion we have:

\begin{theorem} \label{T1}
Let $\left\{(\langle w^2\rangle,
\langle w^3\rangle, \epsilon)\right\}$ be a set of sufficiently
smooth
solutions of (\ref{1.1})--(\ref{1.3}) such that the functions
$\hat{\tau} = \langle w^2\rangle/\epsilon$ satisfy the relations
\begin{equation} \label{H2'}
\frac{\partial\hat\tau}{\partial z}
= 0,\qquad \frac{\partial\hat\tau}{\partial t} = (2\alpha -
\gamma)\left(1 +
a\hat\tau^2N^2\right) + \frac{3}{\delta}.
\end{equation}
Assume that $\kappa = \delta$.
Then operator ${\cal H}^1$ is a sign-invariant of system
(\ref{1.1})--(\ref{1.3}).
\end{theorem}

\begin{proof}
Calculating the time derivative, we
obtain
\begin{equation} \label{H3}
\frac{\partial}{\partial t}{\cal H}^1 =
\frac{\partial\langle w^3 \rangle}{\partial t} +
\delta\frac{\partial}{\partial t}\left[\hat\tau\langle w^2\rangle
\frac{\partial\langle w^3\rangle}{\partial z}\right ].
\end{equation}
Using equation (\ref{1.2})
and the assumption that $\partial\hat\tau/\partial z = 0$,
we can rewrite (\ref{H3}) as follows
\begin{gather*}
\frac{\partial}{\partial t}{\cal H} =
\frac{\partial}{\partial z}\left[\kappa\hat\tau
\langle w^2\rangle\frac{\partial\langle
w^3\rangle}{\partial z} \right] - 3\langle
w^2\rangle\frac{\partial\langle w^2\rangle}{\partial z}
- \gamma\frac{\langle w^3\rangle}{\hat\tau}\left(1 +
a\hat\tau^2N^2\right)\\
\phantom{\frac{\partial}{\partial t}{\cal H} =
}{}+\delta\frac{\partial\hat\tau}
{\partial t}\frac{\partial\langle
w^2\rangle}{\partial z}\langle w^2\rangle
+ \delta\hat\tau\frac{\partial\langle w^2\rangle}{\partial t}
\frac{\partial\langle w^2\rangle}{\partial z}
 + \delta\hat\tau\langle w^2\rangle\frac{\partial^2\langle
 w^2\rangle}
{\partial t\partial z}.
\end{gather*}
Replacing the derivatives $\partial\langle w^2\rangle/\partial t$
and
$\partial^2\langle w^2\rangle/\partial t\partial z$ by their
representations from equation (\ref{1.1}) (taking into account
that $\partial \hat\tau/\partial z = 0$), we have
\begin{gather*}
\frac{\partial}{\partial t}{\cal H}^1 =
\kappa\hat\tau\frac{\partial\langle w^2\rangle}{\partial z}
\frac{\partial\langle w^3\rangle}{\partial z} +
\kappa\hat\tau\langle w^2\rangle\frac{\partial^2\langle
w^3\rangle}
{\partial z^2} - 3\langle w^2\rangle\frac{\partial\langle
w^2\rangle}
{\partial z}
- \gamma\frac{\langle w^3\rangle}{\hat\tau}\left(1 +
a\hat\tau^2N^2\right)\\
\phantom{\frac{\partial}{\partial t}{\cal H}^1 =}{}
+\delta\frac{\partial\hat\tau}{\partial t}\frac{\partial\langle
w^2\rangle}{\partial z}\langle w^2\rangle
+ \delta\hat\tau\left[ -
\frac{\partial\langle w^3\rangle}{\partial z} - \alpha\frac{
\langle w^2\rangle}{\hat\tau}\left(1 +
a\hat\tau^2N^2\right)\right]
\frac{\partial\langle w^2\rangle}{\partial z} \\
\phantom{\frac{\partial}{\partial t}{\cal H}^1 =}{}
+ \delta\hat\tau\langle w^2\rangle\left[ -
\frac{\partial^2\langle w^3\rangle}{\partial z^2} -
\frac{\alpha}{\hat\tau}
\frac{\partial\langle w^2\rangle}{\partial z}\left(1 +
a\hat\tau^2N^2\right)\right].
\end{gather*}
This formula can be written in the form
\begin{gather*}
\frac{\partial}{\partial t}{\cal H}^1 =
\kappa\hat\tau\frac{\partial\langle w^2\rangle}{\partial z}
\frac{\partial\langle w^3\rangle}{\partial z} +
\kappa\hat\tau\langle w^2\rangle\frac{\partial^2\langle
w^3\rangle}
{\partial z^2} - 3\langle w^2\rangle\frac{\partial\langle
w^2\rangle}
{\partial z}
\\
\phantom{\frac{\partial}{\partial t}{\cal H}^1 =}{} -
\frac{\gamma}{\hat\tau}{\cal H}^1\left(1 + a\hat\tau^2N^2\right)
+
\gamma\delta(1 + a\hat\tau^2N^2)\langle
w^2\rangle\frac{\partial\langle w^2\rangle} {\partial z} +
\delta\frac{\partial\hat\tau}{\partial t}\frac{\partial\langle
w^2\rangle}{\partial z}\langle w^2\rangle
\\
\phantom{\frac{\partial}{\partial t}{\cal H}^1 =}{}
- \delta\hat\tau\frac{\partial\langle w^2\rangle}
{\partial z}\frac{\partial\langle w^3\rangle}{\partial z} -
2\delta\alpha\left(1 + a\tau^2N^2\right)
\frac{\partial\langle w^2\rangle}{\partial z}\langle w^2\rangle
- \delta\hat\tau\frac{\partial\langle w^2\rangle}{\partial z}
\frac{\partial^2\langle w^3\rangle}{\partial z^2}.
\end{gather*}
As a result, we obtain
\begin{gather}
\frac{\partial}{\partial t}{\cal H}^1 =
(\kappa - \delta)\hat\tau\frac{\partial\langle
w^2\rangle}{\partial z}
\frac{\partial\langle w^3\rangle}{\partial z} +
(\kappa - \delta)\hat\tau\langle w^2\rangle
\frac{\partial^2\langle w^3\rangle}{\partial z^2} +\langle
w^2\rangle\frac{\partial\langle w^2\rangle}
{\partial z}\Big(\delta\frac{\partial\hat\tau}{\partial
t}\nonumber\\
\phantom{\frac{\partial}{\partial t}{\cal H}^1 = } {} -
2\alpha\delta\left(1 + a\hat\tau^2N^2\right) +
\gamma\delta\left(1 +
a\hat\tau^2N^2\right) - 3\Big) -
\frac{\gamma}{\hat\tau}\left(1 + a\hat\tau^2N^2\right){\cal H}^1.
\end{gather}
It follows from the equality $\kappa = \delta$ and (\ref{H2'})
that
\[
\frac{\partial}{\partial t}{\cal H}^1 =
\frac{\gamma}{\tau} \left(1 + a\hat\tau^2N^2\right){\cal H}^1.
\]
Therefore
\[
\left.{\cal H}^1\right|_{t = t_1} =
\left.{\cal H}^1\right|_{t =
t_0}\exp\left(-\gamma\int_{t_0}^{t_1}\left(\frac{1}
{\hat\tau} + a\hat\tau N^2\right)ds\right).
\]
This completes the proof of the theorem.
\end{proof}

\begin{theorem} \label{T2}
Let $a(2\alpha - \gamma) = \alpha(\varrho - a)$,
$\frac{3}{\delta} + 2\alpha - \gamma = \varrho - \alpha$
and $\kappa = \delta$. Then system (\ref{1.1})--(\ref{1.3})
admits
the invariant set $D$ and its reduction on the set $D$
is of the form:
\begin{gather} \label{H4}
\langle w^2\rangle =
\hat\tau\epsilon,
\\ \label{H5}
\langle w^3\rangle =
-\delta\hat\tau\langle w^2\rangle \frac{\partial\langle
w^2\rangle}{\partial
z},
\\ \label{H6}
\frac{\partial\epsilon}{\partial t}  =  \frac{\partial}
{\partial z}\left[\delta\hat\tau
\langle w^2\rangle\frac{\partial\epsilon}{\partial z}\right]
 - \varrho\left(1 +
 a\hat\tau^2N^2\right)\frac{\epsilon}{\hat\tau},
\end{gather}
where the function $\hat\tau(z,t) \equiv \hat{\tau}(t)$ solves
the ordinary differential equation
\begin{equation} \label{H7}
\frac{d\hat\tau}{dt} = (\varrho - \alpha)\left(1 +
a\hat\tau^2N^2\right)\qquad (\text{{\it a version of equation} }
(\ref{7})
 \text{ {\it on the set }} D).
\end{equation}
\end{theorem}

\begin{proof}
Observe that, according to Theorem~\ref{T1} the invariant set
exists if the equation for~$\hat\tau$
admits a class of solutions which satisfy conditions~(\ref{H2'}).
To find this class of solutions,
we consider equation (\ref{7}) for~$\hat\tau$.
As we have indicated above, this equation is a consequence
of equations (\ref{1.1}), (\ref{1.3}).
More precisely, calculating the
time derivative for $\hat{\tau}$, we obtain
\begin{gather*}
\frac{\partial\hat\tau}{\partial t}
= \frac{1}{\epsilon}
\frac{\partial\langle w^2\rangle}{\partial t} - \frac{\langle
w^2\rangle}{\epsilon^2}\frac{\partial\epsilon}{\partial t}
= - \frac{1}{\epsilon}\frac{\partial\langle w^3\rangle}{\partial
z} -\alpha\left(1 + a\hat\tau^2N^2\right) -
\delta\hat\tau^2\frac{\partial^2\langle
w^2\rangle}{\partial z^2} \\
\phantom{\frac{\partial\hat\tau}{\partial t}
=}{}+ \delta\langle
w^2\rangle\hat\tau\frac{\partial^2\hat\tau}{\partial z^2} -
\delta\frac{\hat\tau^2}{\langle
w^2\rangle}\left(\frac{\partial\langle
w^2\rangle}
{\partial z}\right)^2 + \varrho\left(1 + a\hat\tau^2N^2\right) +
2\delta\frac{\partial\hat\tau}{\partial z}\frac{\partial\langle
w^2\rangle}
{\partial z} -
\delta\frac{\partial\hat\tau}{\partial z}\langle w^2\rangle.
\end{gather*}
Thus the equation for $\hat\tau$ is of the form
\begin{gather*}
\frac{\partial\hat\tau}{\partial t} =
- \frac{\hat\tau}{\langle w^2 \rangle}
\left[\frac{\partial\langle w^3\rangle}{\partial z}
+\delta\hat\tau\langle w^2\rangle\frac{\partial^2\langle
w^2\rangle}{\partial z^2} + \delta\hat\tau
\left(\frac{\partial\langle w^2\rangle}{\partial
z}\right)^2\right]
\\
\phantom{\frac{\partial\hat\tau}{\partial t} = }{}
 + \delta\langle w^2\rangle\hat\tau\frac{\partial^2
 \hat\tau}{\partial
z^2}  + 2\delta\hat\tau\frac{\partial\hat\tau}{\partial
z}\frac{\partial\langle w^2
\rangle}{\partial z} -
\delta\left(\frac{\partial\hat\tau}{\partial z}\right)^2\langle
w^2\rangle +
(\varrho - \alpha)\left(1 + a\hat\tau^2N^2\right).
\end{gather*}
Obviously, it is sufficient to check the conditions of
Theorem~\ref{T1}
only for
$(\langle w^2\rangle, \langle w^3\rangle, \hat\tau) \in D$.
It is clear that equation (\ref{7}) on the set $D$ can be
rewritten in the form
\[
\frac{d\hat\tau}{dt} = (\varrho - \alpha)
\left(1 + a\hat\tau^2 N^2\right)
\]
for $\hat{\tau}(z,t) \equiv \hat{\tau}(t)$
(the expression in square brackets equals zero).
Using the equalities $a(2\alpha - \gamma) = \alpha(\varrho - a)$,
$\frac{3}{\delta} + 2\alpha - \gamma = \varrho - \alpha$, we
conclude that the above-mentioned equation and equation defined
by~(\ref{H2'}) coincide with each other.
The proof is completed by
simple checking that the function
$\langle w^2(z,t)\rangle = \hat{\tau}(t)\epsilon(z,t)$ satisfies
identically~(\ref{1.1}) on the set $D$, where $\hat{\tau}$ solves
the
equations~(\ref{H7}). \end{proof}

Theorem~\ref{T1} is of a special interest in view of its
application to
Turbulent Models.
\begin{corollary} \label{C2}
The equation ${\cal H}^1(\langle w^2\rangle, \langle w^3\rangle,
\hat\tau)$ $=0$ that defines an invariant set of
(\ref{1.1})--(\ref{1.3}) coincides with the algebraic triple
correlation model or the Zeman--Lumley model~\cite{ZEM}.
\end{corollary}

In other words, the algebraic expression represents the equation
of an
invariant set (manifold) generated by the differential equation
for the triple
correlation.

\section{Solutions on invariant sets}

Theorem\,\ref{T2} enables us to reduce (\ref{1.1})--(\ref{1.3})
to the algebraic differential expressions
(\ref{H4})--(\ref{H7}) which can be easier analyzed.
Using the obtained reduction, we construct explicit solutions to
system (\ref{1.1})--(\ref{1.3}) for $N^2\neq 0$. In the case of
nonstratified flow,
i.e.\ for $N^2 \equiv 0$, it was proven
in \cite{GI,IG} that system (\ref{1.1})--(\ref{1.3})
admits a
parametric group of scale
transformation that enables us to find a selfsimilar solution of
the form
\begin{gather*}
\epsilon_a   = \frac{ h(\xi)}{(t+t_0)^{3\mu+\nu}}, \qquad
\langle w^2_a\rangle = \frac{f(\xi)}{(t+t_0)^{2\mu}}, \qquad
\langle w^3_a\rangle =  \frac{q(\xi)}{(t+t_0)^{3\mu}},
\\
\xi  =  \frac{z - z_c}{L}, \qquad
L = \lambda(t+t_0)^{\nu}, \qquad
z_c = \lambda_0L + \lambda_1, \qquad t_0 >0,
\end{gather*}
where $\lambda$, $\lambda_i$ are model constants, $t_0$ is a
parameter and $\nu = 1 - \mu$. In \cite{GI,IG} we studied
the existence
of selfsimilar solutions and presented in detail the qualitative
properties of the
solution obtained.

A direct calculation shows that
for stratified flows there are no selfsimilar solutions
similar to $\epsilon_a$, $\langle w^2_a\rangle$, $\langle
w^3_a\rangle$.
Nevertheless we find a class of explicit solutions
(\ref{1.1})--(\ref{1.3}) by using the fact that
system (\ref{1.1})--(\ref{1.3}) is equivalent
to algebraic differential expressions (\ref{H4})--(\ref{H7})
under the above-mentioned hypotheses on the parameters of the
model.

Let us rewrite (\ref{H7}) in the form
\begin{equation} \label{H8}
\frac{d\hat\tau}{dt} =  BN^2\hat\tau^2(t) + D,
\end{equation}
where $B = a(2\alpha - \gamma)$,
$D = 3/\delta + 2\alpha - \gamma$. We note that
the form of solutions to (\ref{H8}) depends on the signs of
quantities $B$, $D$ and $N^2$. Positiveness (negativeness) of
$N^2$
corresponds to the case of stable (unstable)
stratification.  The signs $B$ and $D$ are determined by values
of coefficients $c_{**}$ which are experimentally known
numbers. The calculated values of numbers $B$ and
$D$ show that $B < 0$, $D > 0$.

\subsection{Stable stratification}
Integrating (\ref{H8}) and denoting by
\[
A = \sqrt{\frac{D}{-BN^2}}
\]
we can obtain the positive solution $\hat\tau_s = \hat\tau_s(t)$
which is defined by the formula
\[
\hat\tau_s(t) = A\tanh\left(4\sqrt{(D/-B)}BNt + C_0\right),
\]
where $C_0$ is a constant. Given the initial data
$\hat\tau_s(t_0)$, the constant $C_0$ can be easily
determined.
We note that function $\hat\tau_s(t)$ has a horizontal
asymptote;
more exactly, $\hat\tau_s(t)\to A$ as $t\to\infty$.

Our main aim is to find a solution to (\ref{H6}).
The initial condition for equation (\ref{H6}) is determined by
the physical model. We have
\[ \epsilon_0(z) \equiv \epsilon(z,0) = \left\{
\begin{array}{ll}
\epsilon_-, & \mbox{if } z < 0,\\
\epsilon_+, & \mbox{if } z\geq 0,
\end{array}
\right. \]
where $\epsilon_-$, $\epsilon_+$ are the positive numbers such
that $\epsilon_- \ne \epsilon_+$ and we assume that
$\epsilon_- < \epsilon_+$.
Set
\[
\theta\equiv\theta(t) = \int_0^t\hat\tau^2_s(p)dp,\qquad
\varsigma(\theta) =
\hat\tau^2_s\left(\theta^{-1}(t)\right),\qquad
\psi(\theta) = \frac{1 +
a\varsigma^2(\theta)N^2}{\varsigma^3(\theta)}
\]
and
\[
\hat\epsilon(z,\theta) =
u_s(z,\theta)\exp\left(-\int_0^{\theta}\psi(p)dp\right),
\qquad\mbox{where} \quad
\hat\epsilon(z,\theta) = \epsilon(z,t).
\]
The function $\theta(t)$ maps $[0,+\infty)$
onto $[0,+\infty)$ and for $u_s$ we have:
\begin{gather} \label{2.1}
\frac{\partial u_s}{\partial\hat\theta}  =
\frac{\partial} {\partial z}\left[\delta u_s \frac{\partial
u_s}{\partial z}\right],\qquad\mbox{where}\quad \hat\theta =
\int_0^\theta\exp\left(-\int_0^\xi\psi(p)dp\right)d\xi,
\\ \label{2.2}
u_s(z, 0) = \epsilon_- \quad \mbox{if} \ \ z < 0, \qquad
u_s(z, 0) = \epsilon_+ \quad \mbox{if} \ \ z \geq 0.
\end{gather}
It is easy to check that $\hat\theta\,:\, [0,+\infty)\to
[0,\hat\theta_0)$, where $\hat\theta_0 < A/aN^2$ and
\[
\exp\left(-\int_0^{\theta}\psi(p)dp\right)\to 0\qquad\mbox{as}
\quad \theta\to
+\infty.
\]
In studying the Cauchy problem for equation (\ref{H6}), we base
our analysis on investigation of transformed problem (\ref{2.1}),
(\ref{2.2}) for a finite time interval $[0,\hat\theta_0)$.
Equation (\ref{2.1}) is usually
called the porous medium equation. It is well-known that
solutions to (\ref{2.1}), (\ref{2.2}) are unique and invariant
under a parametric group of the scale transformation
(see~\cite{VAZ}).
Therefore $u_s$ is a selfsimilar solution which can
be represented in the form
\[
u_s(z,\hat\theta) =
u_s(\hat\xi),\qquad \hat\xi = \frac{z - z_c}{\sqrt{2\hat\theta}},
\qquad z_c(\hat\theta) = \lambda_0\sqrt{2\hat\theta},
\]
where $\lambda_0$ is a model constant, and then (\ref{2.1}),
(\ref{2.2})
is rewritten as
\begin{gather} \label{2.3}
2\delta\frac{d^2u_s}{d\hat\xi^2}  + 2\delta\left(\frac{du_s}
{d\hat\xi}\right)^2  + (\hat\xi + \lambda_0)\frac{du_s}{d\hat\xi}
= 0,
\\ \label{2.4}
u_s(-\infty) = \epsilon_-,\qquad u_s(+\infty) = \epsilon_+.
\end{gather}
Here $z_c(\hat\theta)$ is the so-called central line.
Setting $u_s(\hat\xi) = d\hat\xi/d\zeta$ where $\zeta$ is a new
variable, we obtain the following boundary value problem for the
Blasius equation~\cite{KOK}
\begin{gather} \label{2.5}
2\delta\frac{d^3\hat\xi}{d\zeta^3}  +
(\hat\xi + \lambda_0)\frac{d^2\hat\xi}{d\zeta^2} = 0,
\\  \label{2.6}
\frac{d\hat\xi}{d\zeta}\Big|_{-\infty} =
\epsilon_-,\qquad
\frac{d\hat\xi}{d\zeta}\Big|_{+\infty} = \epsilon_+.
\end{gather}
Equation (\ref{2.5}) arises in the context of the evolution of
boundary layers in an incompressible fluid along a
surface~\cite{KOK}.
As a result, we have that there exists an
one-parametric
family of solutions to (\ref{2.5}), (\ref{2.6}). This implies
existence
of a solution to (\ref{2.3}), (\ref{2.4}).
In~\cite{GI} we proven this result directly for a problem of type
(\ref{2.3}), (\ref{2.4}) and noted that this solution is
essentially different
from the well-known Barenblatt's solution~\cite{BAR}.

\begin{remark} \label{R1}
For the first time, the class of positive solutions to the porous
medium equation
was introduce in~\cite{KOC}.
\end{remark}

Thus we arrive at the following
\begin{lemma} \label{L1}
For any positive finite $\epsilon_-$ and $\epsilon_+$
($\epsilon_- < \epsilon_+$)
there exists a unique positive solution $u_s$ to (\ref{2.3}),
(\ref{2.4})
(respectively (\ref{2.1}), (\ref{2.2})) such that function
$u_s$ is increasing
over ($-\infty,+\infty$); moreover $u_s$ has convex and concave
profiles
for $\hat\xi < 0$ and
$\hat\xi > 0$ respectively.
\end{lemma}

\begin{remark} \label{R2}
The test configurations of profiles of the spectral flux
$\epsilon $ obtained by numerical and experimental methods (see,
for example \cite{KO,VEE}) coincides qualitatively
with profiles of $u_s $.
\end{remark}

Once we have determined $u_s$, we can find
$\langle w^2\rangle$. The function
$\langle w^2\rangle$  is defined from the relation
\[
\langle\hat w^2(z,\theta)\rangle  =
\hat\tau_s\hat\epsilon(z,\theta) =
\hat\tau_su_s(z,\theta)
\exp\left(-\int_0^{\theta}\psi(p)dp\right), \qquad \langle
w^2(z,t)\rangle =
\langle\hat w^2(z,\theta)\rangle.
\]
For the triple correlation
$\langle w^3\rangle$ we obtain
\begin{equation}\label{2.7}
\langle\hat w^3(z,\theta)\rangle =
-\delta\hat\tau_s
\langle\hat w^2(z,\theta)\rangle\frac{\partial\langle\hat
w^2(z,\theta)\rangle}{\partial z}, \qquad
\langle w^3(z,t)\rangle =
\langle\hat w^3(z,\theta)\rangle.
\end{equation}

\subsection{Unstable stratification}

Let us now find a solution to (\ref{H8}) for $N^2 < 0$.
In this case we arrive at the following solution to equation
(\ref{H8})
\[
\hat\tau_{ns} = A\tan\left(\sqrt{(DBN^2)}t + C_1\right),
\]
where $C_1$ is determined by the initial data $\hat\tau(0)$.
As before, we introduce into consideration the new time
variable
\[
\theta = \theta(t) \equiv \int_0^t\hat\tau^2_{ns}(p)dp
\]
and note that $\theta\to +\infty$ as $\sqrt{DBN^2}t +
C_1\to\pi/2$.
Then we obtain
\begin{gather*}
u_{ns}(z,\theta) =
\hat\epsilon(z,\theta)\exp\left(\int_0^{\theta}
\psi(p)dp\right),\\
\psi(\theta) = \frac{1 +
a\varsigma^2(\theta)N^2}{\varsigma^3(\theta)},\qquad
\varsigma(\theta) = \hat\tau^2_{ns}\left(\theta^{-1}(t)\right),
\end{gather*}
and
\[
\frac{\partial u_{ns}}{\partial\hat\theta} = \frac{\partial}
{\partial z}\left[\delta u_{ns}
\frac{\partial u_{ns}}{\partial z}\right],\qquad\mbox{where}\quad
\hat\theta =
\int_0^\theta\exp\left(-\int_0^\xi\psi(p)dp\right)d\xi.
\]
Here $\hat\theta \to +\infty$ as $\sqrt{(DBN^2)}t +
C_1\to\pi/2$
and $\exp\left(-\int_0^{\theta}\psi(s)ds\right)$ is an increasing
bounded
function on
the interval $[0,+\infty)$ that coincides with experimental
observations about
increasing the spectral flux $\epsilon$ in the case of unstable
stratification
of the flow. Further analysis goes along the same lines just as
for $N^2 > 0$.
Combining Lemma~\ref{L1} with the above arguments, we claim
the following

\begin{theorem} \label{T3}
Let $\kappa$ = $\delta$ and
\[
a(2\alpha - \gamma) = \alpha(\varrho - a),\qquad
\frac{3}{\delta} + 2\alpha - \gamma = \varrho - \alpha.
\]
Then there exists a solution to system (\ref{1.1})--(\ref{1.3})
of the following form
\begin{gather*}
\hat\epsilon(z,\theta) =
u_s(z,\theta)\exp\left(\!-\int_0^{\theta}\psi(s)ds\right),
\quad \theta = \theta(t) \equiv \int_0^t\hat\tau^2_s(p)dp,\quad
\epsilon(z,t) = \hat\epsilon(z,\theta),
\\
\langle w^2(z,t)\rangle  = \hat\tau_s\epsilon(z,t),\qquad
\langle w^3(z,t)\rangle =
-\delta\hat\tau_s
\langle w^2(z,t)\rangle\frac{\partial\langle
w^2(z,t)\rangle}{\partial z}\qquad \text{{\it for}} \quad  N^2>0,
\end{gather*}
and
\begin{gather*}
\hat\epsilon(z,\theta) =
u_{ns}(z,\theta)\exp\left(\!-\int_0^{\theta}\psi(s)ds\right),
\quad \theta = \theta(t) \equiv
\int_0^t\hat\tau^2_{ns}(p)dp,\quad
\epsilon(z,t) = \hat\epsilon(z,\theta),
\\
\langle w^2(z,t)\rangle  = \hat\tau_{ns}\epsilon(z,t),\qquad
\langle w^3(z,t)\rangle =
-\delta\hat\tau_{ns}
\langle w^2(z,t)\rangle\frac{\partial\langle
w^2(z,t)\rangle}{\partial z}\qquad \text{{\it for}} \quad N^2 <
0,
\end{gather*}
where $u_s\,(u_{ns})$ is a selfsimilar solution to (\ref{2.3}),
(\ref{2.4}).
\end{theorem}

\section{Conclusions}

As a rule, the parametric models of turbulence represent the
transport equations for the second-order moments which are
completed by closure relations for the higher-order moments and
dissipation tensor, and in many cases these closure relations
are given in the so-called isotropic form. The assumption about
the relaxation character of turbulence under its evolution to
the equilibrium state (the homogeneous isotropic state with
Gaussian distribution of turbulent fluctuations) is the base
for such simplification. However, these models describe
adequately certain statistical structures of investigated flows
even if turbulence is characterized by anisotropic effects. In
general, conditions (\ref{H2'}) (which assume ``uniformity''
of the turbulent scales) can be used in the case of an
equilibrium state of turbulent flows. Nevertheless, the
results obtained provide a correct modelling for turbulent flows:
the algebraic model for triple correlations (\ref{H5}) was
used in~\cite{ZEM} for studying turbulent structures in the
convective boundary layer and moreover, it was testified for
distinct turbulent flows (see~\cite{IL1}).

We conclude with some observations and comments.
A version of formula
(\ref{H5}) coincides with the well-known Hanjalic--Launder
model~\cite{HAN} in the case of nonstratified flow~\cite{GI}.
It is possible to get results similar to the results of this
article for shear flows in the problem of plane turbulent wake.
Maybe the most interesting and novel part of the criterion of
invariance for the so-called locally equilibrium approximations
(see~\cite{LH})
in the problem of plane turbulent wake consists in showing that
the Poisson bracket $\{U,e\}$ equals zero. Here $U$ is the
velocity excess and $e$ is the turbulent kinetic energy.
An example of such an equality was obtained in~\cite{GR} for
a selfsimilar solution.

\subsection*{Acknowledgements}

This research was partially supported by Integration
Project SD RAS (grant no.~2000-01), RFBR (grant no.~01-01-00783).
This work
was supported by INTAS (proposal no. 97-2022).

\label{grebenev-lastpage}

\end{document}